\documentclass{eptcs}
\usepackage{breakurl}
\usepackage{amsmath, amssymb, latexsym}
\usepackage{graphicx}

\title{Imitation in Large Games}
\author{Soumya Paul
\institute{The Institute of Mathematical Sciences\\
Chennai, India - 600~113}
\email{soumya@imsc.res.in}
\and
R.~Ramanujam
\institute{The Institute of Mathematical Sciences\\
Chennai, India - 600~113}
\email{jam@imsc.res.in}}
\def\titlerunning{Imitation in Large Games}
\def\authorrunning{Soumya Paul \& R. Ramanujam}
\begin{document}

\maketitle


\newcommand{\players}{\ensuremath{N}}
\newcommand{\actions}{\ensuremath{A}}
\newcommand{\Atuple}{\ensuremath{\textbf{A}}}
\newcommand{\atuple}{\ensuremath{\textbf{a}}}
\newcommand{\tree}{\ensuremath{\mathcal{T}}}
\newcommand{\tarrow}{\ensuremath{\Rightarrow}}
\newcommand{\nodes}{\ensuremath{T}}
\newcommand{\node}{\ensuremath{t}}
\newcommand{\hist}{\ensuremath{t}}
\newcommand{\tstrat}{\ensuremath{\mu}}
\newcommand{\tStrat}{\ensuremath{\Omega}}
\newcommand{\pstrat}{\ensuremath{\sigma}}
\newcommand{\pStrat}{\ensuremath{\Sigma}}
\newcommand{\strat}{\ensuremath{\tau}}
\newcommand{\stree}{\ensuremath{\widetilde{\mathcal{T}}}}
\newcommand{\tp}{\ensuremath{\mathcal{TP}}}
\newcommand{\pt}{\ensuremath{\mathcal{PT}}}
\newcommand{\prop}{\ensuremath{\mathcal{P}}}
\newcommand{\val}{\ensuremath{\mathit{Val}}}
\newcommand{\chop}{\ensuremath{\smallfrown}}
\newcommand{\oStrat}{\ensuremath{\Pi}}
\newcommand{\ostrat}{\ensuremath{\pi}}
\newcommand{\istrat}{\ensuremath{\gamma}}
\newcommand{\iStrat}{\ensuremath{\Gamma}}
\newcommand{\otest}{\ensuremath{\psi}}
\newcommand{\itest}{\ensuremath{\varphi}}
\newcommand{\false}{\ensuremath{\bottom}}
\newcommand{\true}{\ensuremath{\top}}
\newcommand{\sembrack}[1]{[\![#1]\!]}
\newcommand{\io}{\ensuremath{h}}
\newcommand{\oi}{\ensuremath{g}}
\newcommand{\arena}{\ensuremath{\mathcal{G}}}
\newcommand{\game}{\ensuremath{\mathcal{G}}}
\newcommand{\vertices}{\ensuremath{W}}
\newcommand{\edges}{\ensuremath{E}}
\newcommand{\vertex}{\ensuremath{w}}
\newcommand{\move}[1]{\ensuremath{\stackrel{#1}{\rightarrow}}}
\newcommand{\succr}[1]{\ensuremath{{#1}_\rightarrow}}
\newcommand{\fst}{\ensuremath{\mathcal{A}}}
\newcommand{\run}{\ensuremath{\chi}}
\newcommand{\lang}{\ensuremath{\mathcal{L}}}
\newcommand{\qed}{\hfill\ensuremath{\Box}}
\newcommand{\diamondminus}{\ensuremath{\Diamond\!\!\!\!\!\! -}}
\newcommand{\start}{\ensuremath{\triangleright}}
\newcommand{\avertices}{\ensuremath{V}}
\newcommand{\agraph}{\ensuremath{G}}
\newcommand{\averts}{\ensuremath{V}}
\newcommand{\aedges}{\ensuremath{E}}
\newcommand{\avertex}{\ensuremath{v}}
\newcommand{\gtest}{\ensuremath{\phi}}
\newcommand{\Gtest}{\ensuremath{\Phi}}
\newcommand{\sarena}{\ensuremath{Z}}
\newcommand{\svertex}{\ensuremath{z}}
\newcommand{\flc}{\ensuremath{\mathit{CL}}}
\newcommand{\md}{\ensuremath{\mathit{md}}}
\newcommand{\bigo}{\ensuremath{\mathcal{O}}}
\newcommand{\rest}{\ensuremath{\!\upharpoonright\!}}
\newcommand{\mbound}{\ensuremath{r}}
\newcommand{\amap}{\ensuremath{h}}
\newcommand{\omap}{\ensuremath{g}}
\newcommand{\testset}{\ensuremath{S}}
\newcommand{\upd}{\ensuremath{\mathit{Update}}}
\newcommand{\safe}{\ensuremath{S}}
\newcommand{\muller}{\ensuremath{\mathcal{F}}}
\newcommand{\attr}{\ensuremath{\mathit{Attr}}}
\newcommand{\rank}{\ensuremath{r}}
\newcommand{\bari}{\ensuremath{{\bar\imath}}}
\newcommand{\nxt}{\ensuremath{\mathit{next}}}
\newcommand{\fintree}{\ensuremath{\mathcal{T}_\mathit{LAR}}}
\newcommand{\sfintree}{\ensuremath{\mathcal{T}^*_\mathit{LAR}}}
\newcommand{\bc}{\ensuremath{\mathit{BC}}}
\newcommand{\cupover}{\ensuremath{\bigcup}}
\newcommand{\step}[1]{\ensuremath{\stackrel{#1}{\rightarrow}}}
\newcommand{\R}{\ensuremath{\mathcal{R}}}

\newtheorem{proposition}{Proposition}
\newtheorem{definition}{Definition}
\newtheorem{lemma}{Lemma}
\newtheorem{theorem}{Theorem}
\newtheorem{claim}{Claim}
\newtheorem{procedure}{Procedure}[subsection]
\newenvironment{reptheorem}[1]{\noindent {\bf Theorem~#1.~}}{\smallskip}

\newenvironment{proof}{\vspace{1ex}\noindent{\bf
Proof}\hspace{0.5em}}{\hfill\qed\vspace{1ex}}
\newenvironment{pfoutline}{\vspace{1ex}\noindent{\bf Proof
 Outline}\hspace{0.5em}}{\hfill\qed\vspace{1ex}}
\newenvironment{rem}{\vspace{1ex}\noindent{\bf
Remark}\hspace{0.5em}}{\hfill\vspace{1ex}}


\begin{abstract}
In games with a large number of players where players may have
overlapping objectives, the analysis of stable outcomes typically depends on
player types. A special case is when a large part of
the player population consists of imitation types: that of players who
imitate choice of other (optimizing) types. Game theorists
typically study the evolution of such games in dynamical systems
with imitation rules.  In the setting of games of infinite duration on
finite graphs with preference orderings on outcomes for player types, we
explore the possibility of imitation as a viable strategy. In our setup, the
optimising players play bounded memory strategies and the imitators play
according to specifications given by automata. We present algorithmic
results on the eventual survival of types.
\end{abstract}


\section{Summary}\label{sec1}
Imitation is an important heuristic studied by game theorists in
the analysis of large games, in both extensive form games with 
considerable structure, and repeated normal form games with large 
number of players. One reason for this is that notions of rationality
underlying solution concepts are justified by players' assumptions
about how other players play, iteratively. In such situations,
players' knowledge of the types of other players alters game dynamics.
Skilled players can then be imitated by less skilled ones, and 
the former can then strategize about how the latter might play.
In games with a large number of players, both strategies and outcomes
are studied using distributions of player types.

The dynamics of imitation, and strategizing of optimizers in the
presence of imitators can give rise to interesting consequences.
For instance, in the game of chess, if the player playing white somehow 
knows that her opponent will copy her move for move then the following 
simple sequence of moves allows her to checkmate her opponent 
\footnote{This is called `monkey-chess' in chess parlance.}:

\texttt{1.e3 e6 2.Qf3 Qf6 3.Qg3 Qg6 4.Nf3 Nf6 5.Kd1 Kd8 6.Be2
  Be7 7.Re1 Re8 \linebreak[4]{8.Nc3 Nc6} 9.Nb5 Nb4 10.Qxc7\#}

On the other hand, we can have the scenario where every player is
imitating someone or the other and the equilibrium attained maybe 
highly inefficient.  This is usually referred to as `herd behaviour' 
and has been studied for instance in \cite{Ban92}.

In an ideal world, where players have unbounded resources and
computational ability, each of them can compute their optimal strategies
and play accordingly and thus we can predict optimal play. But in
reality, this is seldom the case. Players are limited in their
resources, in computational ability and their knowledge of the game. 
Hence, in large games it is not possible for such players to compute their 
optimal strategies beforehand by considering all possible scenarios that 
may arise during play. Rather, they observe the outcome of the game and then
strategise dynamically.  In such a setting again, imitation types make sense.

A resource bounded player may attach some cost to strategy
selection. For such a player, imitating another player who has been doing
extensive research and computation may well be worthwhile, even if her
own outcomes are less than optimal. What is lost in sub-optimal
outcomes may be gained in avoiding expensive strategisation.

Thus, in a large population of players, where resources and computational 
abilities are asymmetrically distributed, it is natural to consider
a population where the players are predominantly of two kinds: optimisers and
imitators.\footnote{There would also be a third kind of players,
  randomisers, who play any random strategy, but we do not consider
  such players in this exposition.} Asymmetry in resources and abilities
can then lead to different types of imitation and thus ensure that
we do not end up with `herd behaviour' of the kind referred to above.
Mutual reasoning and strategising process between optimizers and
imitators leads to interesting questions for game dynamics in 
these contexts.

Imitation is typically modelled in the dynamical systems framework
in game theory.  Schlag (\cite{Sch98}) studies a model of repeated games where 
a player in every round samples one other player according to some sampling 
procedure and then either imitates this player or sticks to
her own move. He shows that the strategy where a player imitates the
sampled player with a probability that is proportional to the
difference in their payoffs, is the one that attains the maximum
average payoff in the model. He also gives a simple counterexample to
show that the na\"ive strategy of `imitate if better' may not always
be improving. Banerjee (\cite{Ban92}) studies a sequential decision
model where each decision maker may look at the decisions made by the
previous decision makers and imitate them. He shows that the decision
rules that are chosen by optimising individuals are characterised by
herd behaviour, i.e., people do what others are doing rather than
using their own information. He also shows that such an equilibrium is
inefficient. Levine and Pesendorfer (\cite{LP07}) study a model where 
existing strategies are more likely to be imitated than new strategies 
are to be introduced. 

The common framework in all of the above studies is repeated non-zero-sum
normal form games where the questions asked of the model are somewhat different
from standard ones on equilibria. Since all players are not optimizers,
we do not speak of equilibrium profiles as such but optimal strategies
for optimizers and possibly suboptimal outcomes for imitators. In the
case of imitators, since they keep switching (imitate $i$ for 2 moves,
$j$ for 3 moves, then again $i$ for 1 move, etc.) studies consider {\bf 
stability} of imitation patterns, what types of imitation survive
eventually, since these would in turn determine play by optimizers and
thus stable subgames, thus determining stable outcomes.  Note that, as 
in the example of chess above, imitation and hence the study of system 
dynamics of this kind, makes equal sense in large turn based extensive 
form games among resource bounded players as well. 

For finitely presented infinite games the stability questions above
can be easily posed and answered in automata theoretic ways, since
typically bounded memory strategies suffice for optimal solutions,
and stable imitation patterns can be analysed algorithmically. Indeed,
this also provides a natural model for resource bounded players as
finite state automata. 

With this motivation, we consider games of unbounded duration on
finite graphs among players with overlapping objectives where 
the population is divided into players who optimise and others 
who imitate. Unbounded play is natural in the study of imitation
as a heuristic, since `losses' incurred per move may be amortised 
away and need not affect eventual outcomes very much. Imitator
types specify how and who to imitate and are given using finite
state transducers. Since plays eventually settle down to
connected components, players' preferences are given using orderings
on Muller sets \cite{PS09}. In this work, we study turn-based games
so as to use the set of techniques already available for the analysis
of such games.

In this setting we address the following questions and present
algorithmic results:

\begin{itemize}
\item If the optimisers and the imitators play according to certain
  specifications, is a global outcome eventually attained? 
\item What sort of imitative behaviour (subtypes) eventually survive
  in the game?
\item How worse-off are the imitators from an equilibrium outcome?
\end{itemize}

Infinite two-player turn-based games on finite graphs have been
extensively studied in the 
literature. A seminal result by B\"uchi and Landweber
\cite{BL69} showed that for Muller objectives, winning strategies
exist in bounded memory strategies and can be effectively
synthesised. Martin \cite{Mar75} showed that such games with Borel
winning conditions are sure-determined (one of the players always has a
winning strategy from every vertex). Zielonka \cite{Zie98} gave an
insightful analysis of Muller games and provided an elegant algorithm
to compute bounded memory winning strategies.

For concurrent-move games, sure determinacy does not hold, and the
optimal value determinacy (the values of both the players at every
vertex sum to 1) for concurrent-move games with Borel objectives was
proved in \cite{Mar98}. Concurrent games with qualitative reachability
and more general parity objectives have been studied in
\cite{dAHK98,dAH00}. Such games have also been extended to the
multiplayer
setting where the objectives of the players are allowed to
overlap. \cite{CJM04,GU08} show that when the objectives are win-lose
Borel, subgame perfect equilibria
exist. \cite{PS09} show that bounded memory equilibrim tuples exist in
turn based games
even when the objectives are not win-lose but every player has
preferences over the various Muller sets.


\section{Games, Strategies and Objectives}\label{sec2}
The model of games we present is the standard model of turn based games of 
unbounded duration on finite graphs. For any positive integer $n$, let
$[n] =\{1, \ldots, n\}$.

\begin{definition}
Let $n \in \mathbb{N}$, $n > 1$. An $n$-player game arena is a directed graph 
$\game = (V_1, \ldots V_n, A, E)$, where $V_i$ are finite sets of game 
positions with $V_i \cap V_j = \emptyset$ for $i \neq j$, $V =
\cupover_{i \in [n]} V_i$, $A$ is a finite set of moves,
and $E \subseteq (V \times A \times V)$ is the move relation that satisfies the 
following conditions:
\begin{enumerate}
\item For every $v, v_1, v_2 \in V$ and $a, b \in A$, if $(v,a,v_1) \in E$ 
and $(v,b,v_2) \in E$ then $a \neq b$.
\item For every $v \in V$, there exists $a \in A$ and $v' \in V$
such that $(v,a,v') \in E$.
\end{enumerate}
 When an initial position $v_0 \in V$ is specified, we call
 $(\game,v_0)$ an initialised arena or just an arena.
\end{definition}

In this model, we assume for convenience that the moves of all players
are the same.
When $v \in V_i$, we say that player $i$ owns the vertex $v$.
A game arena is thus a finite graph with nodes labelled by players and
edges labelled
by moves such that no two edges out of a vertex share a common label and there
are no dead ends. For a vertex $v\in V$, let $vE$ denote its set of neighbours: 
$vE = \{v' | (v,a,v') \in E$ for some $a \in A \}$. For $v\in V$ and $a\in A$,
let $v[a] = \{v' | (v,a,v') \in E\}$; $v[a]$ is either empty or the singleton
$\{v'\}$. In the latter case, we say $a$ is enabled at $v$ and write
$v[a] = v'$. 
For $u\in A^*$, we can similarly speak of $u$ being enabled at $v$ and define 
$v[u]$ so that when $v[u] = \{v'\}$, there is a path in the graph from $v$ to 
$v'$ such that $u$ is the sequence of move labels of edges along that path.
Given $v\in V$ and $u\in A^*$, if any $u$-labelled path exists in the graph,
it is unique. On the other hand, given any sequence of vertices that correspond 
to a path in the graph, there may be more than one sequence of moves that label 
that path.

A  play in $(\game,v_0)$ is an infinite path $v_0 \step{a_1} \ldots$,
such that $v_i \step{a_{i+1}} v_{i+1}$ for $i \in \mathbb{N}$. We often
speak of $a_0 a_1 \ldots \in A^\omega$ as the play to denote this path.
The game starts by placing a token at $v_0 \in V_i$. Player $i$
chooses an action $a\in A$ enabled at $v_0$ and the token moves along the 
edge labelled $a$ to a neighbouring vertex $v_1 \in V_j$. Player $j$ 
chooses an action $a' \in A$ enabled at $v_1$, the token moves along the edge
labelled $a'$ to a neighbouring vertex and so on.  Note that since there
are no dead ends, any player whose turn it is to move has some available move.  

Given a path $\rho = v \step{a_0} v_1 \ldots \step{a_k} v_k$, we call
  $a_\ell, (1 \leq \ell \leq k)$ the last $i$-move in $\rho$, if $v_{\ell - 1}
  \in V_i$ and for all $\ell': \ell < \ell' \leq k$, $v_{\ell'} \notin V_i$.
\subsection{Objectives}
The game arena describes only legal plays, and the game itself is defined by
specifying outcomes and players' preferences on outcomes. Since each play
results in an outcome for each player, players' preferences are on plays.
This can be specified finitely, as every infinite play on a finite graph
settles down to a strongly connected component.

For a play $u\in A^\omega$ let $\inf(u)$ be the set of vertices that appear 
infinitely often in the play given by $u$. With each player $i$, we associate
a total pre-order $\preceq_i \subseteq (2^V \times 2^V)$. This induces a total
preorder on plays as follows: $u \preceq_i u'$ iff $\inf(u) \preceq_i \inf(u')$.

Thus an $n$-player  game is given by a tuple $(\game, v_0, \preceq_1, \ldots, 
\preceq_n)$, consisting of an $n$-player game arena and players' preferences.

\subsection{Strategies} 
Players strategise to achieve desired outcomes. Formally, a strategy $\sigma_i$ 
for player $i$ is a partial function $$\sigma_i : VA^* \rightharpoonup A$$
where $\sigma_i(vu)$ is defined if $v[u]$ is defined and $v[u] \in V_i$, and
if $(v[u])[\sigma_i(vu)]$ is defined.  

A strategy $\sigma_i$ of player $i$ is said to be bounded memory if
there exists a finite state transducer FST $\fst_\sigma = (M, \delta, g, m_0)$
where $M$ (the `memory' of the strategy) is a finite set of states,
$m_0\in M$ is the initial state of the memory, $\delta:A\times M \rightarrow M$
is the `memory update' function, and $g:V_i\times M\rightarrow A$ is the
`move' function such that for all $v \in V_i$ and $m \in M$,
$g(v,m)$ is enabled at $v$ and the following condition holds:
given $v\in V_i$, when $u = a_1 \ldots a_k \in A^*$ is a partial play 
from $v$, $\sigma_i(vu)$ is defined, $\sigma_i(vu) = g(v[u],m_k)$, where 
$m_k$ is determined by: $m_{i+1}=\delta(a_{i+1},m_i)$ for $0 \leq i < k$.

A strategy is said to be memoryless or positional if $M$ is a singleton. That 
is, the moves depend only on the current position.

\begin{definition}
Given a strategy profile $\bar{\sigma} =
(\sigma_1,\ldots,\sigma_n)$ 
for $n$ players let $\rho_{\bar{\sigma}}$ denote the unique play
in $(\game,v_0)$
conforming to $\bar{\sigma}$. A profile $\bar{\sigma}$ is called a
Nash equilibrium in $(\game,v_0,\prec_1,\ldots,\prec_n)$ if
for every player $i$ and for every 
other strategy 
$\sigma'_i$ of player $i$, $\inf(\rho_{(\bar{\sigma}_{-i},\sigma'_i)})
\preceq_i \inf(\rho_{\bar{\sigma}})$.
\end{definition}


\section{Specification of Strategies}\label{sec3}
We now describe how the strategies of the imitator and optimiser types 
are specified.

\subsection{Imitator Types}
An imitator type is again specified by a finite state transducer which
advises the imitator whom to imitate when using memory states for
switching between imitating one player or another. When deciding not
to imitate any other player, we assume that the type advises what to
play using a memoryless strategy.

An {\sf imitator type} $\tau_j$ for player $j$ is a tuple
$(M, \pi, \mu, \delta, m_0)$ where $M$ is the finite set
denoting the memory of the strategy, $m_0 \in M$ is the initial
memory, $\delta: A\times M\rightarrow M$ is the memory update function,
$\pi: V \rightarrow A$ is a positional strategy such that for any $v
\in V$, $\pi(v)$ is enabled at $v$, and $\mu: M \rightarrow [n]$ 
is the imitation map. 

Given $\tau_j$ as above, define a strategy $\sigma_j$ for player $j$
as follows. Let $v\in V$ and $u = a_1 \ldots a_k \in A^*$ is a partial play
from $v$ such that  $v[u]$ is defined and $v[u] \in V_j$. Let $m_{i+1}=
\delta(a_{i+1},m_i)$ for $0 \leq i < k$. Then $\sigma_j(vu) = a_\ell$,
if $a_\ell$ is the last $\mu(m_k)$ move in the given play and $a_\ell$
is enabled at $vu$, and $\sigma_j(vu) = \pi(v[u])$, otherwise.

Note that the type specification only specifies whom to imitate, and
how it decides whom to imitate but is silent on the rationale for
imitating a player or switching from imitating $x$ to imitating $y$.
In general an imitator would have a set of observables, and based on
observations of game states made during course of play, would decide 
on whom to imitate when. Thus imitator specifications could be given
by a past-time formula in a simple propositional modal logic. With
any such formula we can associate an imitation type transducer as
defined above, so we do not pursue that approach here. See, for
instance, \cite{PRS09} for more along that direction.

The following are some examples of imitating strategies that
can be expressed using such automata:

\begin{enumerate}
\item Imitate player 1 for 3 moves and then keep imitating player 4
  forever.
\item Imitate player 2 till she receives the highest payoff. Otherwise
  switch to imitating player 3.
\item Nondeterministically imitate player 4 or 5 forever.
\end{enumerate}

For convenience of the subsequent technical analysis, we assume that
an imitator type \linebreak[4]{$\tau=(M, \pi, \mu, \delta, m_0)$} is
presented as a
finite state transducer $\R_\tau = (M',\delta',g',m_I)$ where

\begin{itemize}
\item $M' = V\times M\times A^{[n]}$.
\item $\delta': A\times M' \rightarrow M'$ such that
  $\delta'(a,\langle v,m,(a_1,\ldots, a_n)\rangle) = \langle v', m',
  (a_1,\ldots, a_{i-1}, a, a_{i+1}, \ldots, a_n)\rangle$ such that
  $v\step{a} v'$, $\delta(a,m)=m'$ and $v\in V_i$.
\item $g': V\times M' \rightarrow A$ such that $g'(v, \langle
  v,m,(a_1,\ldots, a_n)\rangle) = a_i$ iff $\mu(m)=i$ and $a_i$ is
  enabled at $v$. Otherwise $g'(v, \langle
  v,m,(a_1,\ldots, a_n)\rangle) = \pi(v)$.
\item $m_I = \langle v_0,m_0,(a_1,\ldots,a_n)\rangle$ for some
  $(a_1,\ldots, a_n) \in A^{|n|}$.
\end{itemize}

Figure 1 below depicts an imitator strategy where a player imitates
player 1 for two moves and then player 2 for one move and then again
player 1 for two moves and so on. She just plays the last move of the
player she is currently imitating. Suppose there are a total of $p$
actions, that is, $|A|=p$. She remembers the last move of the
player she is imitating in the states $m_1$ to $m_p$, and when it is
her turn to move, plays the corresponding action.

\begin{figure}[t]
\centering
\includegraphics[scale=0.6]{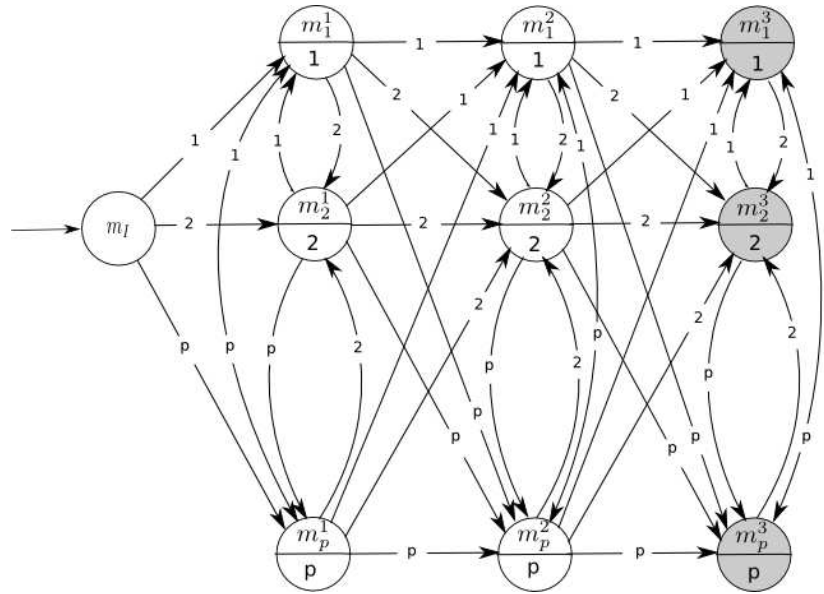}
\label{fig:figure1}
\caption{An imitator strategy}
\end{figure}

Given an FST $\R_\tau$ for an imitator type $\tau$, we call a strongly
connected component of $\R_\tau$ a subtype of $\R_\tau$. We will
often refer to the strategy $\sigma_j$ induced by the imitator type $\R_\tau$
for player $j$ as $\R_\tau$, when the context is clear.

We define the notion of an imitation equilibrium which is a tuple
of strategies for the optimisers such that none of the optimisers can
do better by unilaterally deviating from it given that the imitators
stick to their specifications.
\begin{definition}
In the game $(\game,v_0, \prec_1,\ldots,\prec_n)$, given that the
imitators $r+1,\ldots,n$
play strategies \linebreak[4]{$\tau_{r+1},\ldots,\tau_n$}, a profile
of strategies
$\bar{\sigma}=(\sigma_1,\ldots,\sigma_r)$ of the optimisers is called an
imitation equilibrium if for every optimiser $i$ and for every other strategy
$\sigma'_i$ of  $i$, $\inf(\rho_{(\bar{\sigma}_{-i},\sigma'_i)})
\preceq_i \inf(\rho_{\bar{\sigma}})$.
\end{definition}

\begin{rem}
  Note that an imitation equilibrium $\bar{\sigma}$ may be quite
  different from a Nash equilibrium $\bar{\sigma}'$ of the game
  $(\game,v_0, \prec_1,\ldots,\prec_n)$ when restricted to the first
  $r$ components. In a Nash equilibrium the imitators are not restricted
  to play according to the given specifications unlike in an imitation
  equilibrium. In the latter case, the optimisers, in certain
  situations, may be able to exploit these restrictions imposed on the
  imitators (as in the example of `monkey-chess' discussed in Section
  \ref{sec1}).
\end{rem}

\subsection{Optimiser Specifications}
One of the motivations for an imitator to imitate an optimiser is
the fact that an optimiser plays to get best results. To an imitator,
an optimiser appears to have the necessary resources to compute and
play the best strategy and hence by imitating such a player she cannot
be much worse off. But what kind of strategies do the optimisers
play on their part?

In the next section, we show that if the optimisers know the types
(the FSTs) of each of the 
imitators, then it suffices for them to play bounded memory
strategies. Of course, this depends on the solution concept: Nash
equilibrium is defined 
for strategy profiles, we need to particularize them for applying only to
optimizers.

Thus in the treatment below, we consider only bounded memory strategies for
the optimisers.


\section{Results}\label{sec4}
In this section, we first show that it suffices to consider bounded memory
strategies for the optimisers. Then we go on to address the questions
raised towards the end of Section \ref{sec1}.

First we define a product operation between an arena and a
bounded memory strategy.
\subsection{Product Operation} Let $(\game,v_0)$ be an arena
and $\sigma$ be a bounded memory strategy given by the FST
$\fst_\sigma = (M, \delta, g, m_I)$. We define $\game\times \fst_\sigma$ to
be the graph
$(\game',v'_0)$ where $\game' = (V',E')$ such that
\begin{itemize}
\item $V' = V\times M$
\item $v'_0 = (v_0,m_0)$
\item 
  \begin{itemize}
  \item If $g(v,m)$ is defined then 
    $(v,m)\stackrel{a}{\rightarrow}(v',m')$ iff
    $\delta(a,m)=m'$, $v\stackrel{a}{\rightarrow} v'$ and $g(v,m) =
    a$.
  \item If $g(v,m)$ is not defined then
    $(v,m)\stackrel{a}{\rightarrow}(v',m')$ iff 
    $\delta(a,m)=m'$ and $v\stackrel{a}{\rightarrow} v'$
  \end{itemize}
\end{itemize}

\begin{proposition}\label{prop1}
Let $(\game,v_0)$ be an arena and $\sigma$ be a bounded memory strategy. Then
$\game\times\fst_\sigma$ is an arena, that is, there are no dead ends.
\end{proposition}

\begin{proof} Let $(\game', v'_0) = \game\times\fst_\sigma$. 
$\delta: A\times M\rightarrow M$ being a function, $\delta(a,m)$
is defined for every $a\in A$ and $m\in M$. Also by the definition of
$\game$, for every vertex $v \in V$ there exists an action $a\in A$
enabled at $v$ and a vertex
$v'\in V$ such that $v\step{a} v'$. Thus for every vertex
$(v,m)\in V'$,
\begin{itemize}
\item if $g(v,m)$ is not defined then corresponding to every enabled
  action $a\in A$ there exists
  $(v',m')\in V'$ such that $(v,m)\step{a}(v',m')$,
\item if $g(v,m)$ is defined then by definition the unique action $a=
  g(v,m)$ is enabled at $v$. Hence,
  there exists $(v',m')\in V'$ such that $(v,m) \step{a} (v',m')$.
\end{itemize}
\end{proof}

Thus taking the product of the arena with a bounded memory
strategy $\sigma_i$ of player $i$ does the following. For a vertex
$v\in V_i$, it retains only the outgoing edge that is labelled with the action
specified by the corresponding memory state of $\sigma_i$. For all
other vertices $v\notin V_i$, it
retains all the outgoing edges.

\begin{proposition}
 Let $(\game,v_0)$ be an arena and $\sigma_1,\ldots,\sigma_n$ be bounded
 memory strategies. Then 
$\game\times\fst_{\sigma_1}\times\ldots\times\fst_{\sigma_n}$ is an
arena, that is, there are no dead ends. 
\end{proposition}

\subsection{Equilibrium}
Of the $n$ players let the first $r$ be optimisers and the rest
$n-r$ be imitators. Let
$\tau_{r+1},\ldots,\tau_n$ be the types of the imitators $r+1,
\ldots, n$. We transform the game $(\game,v_0,\prec_1,\ldots,\prec_n)$
with $n$ players to a
game $(\game',v'_0, \prec'_1,\ldots,\prec'_{r+1})$ with  $r+1$ players
in the following steps:
\begin{enumerate}
\item Construct
the graph $(\game',v'_0) = ((V',E'),v'_0)$ as $\game' =
\game\times\R_{\tau_{r+1}}\times\cdots\times\R_{\tau_n}$. 
\item Let $V' = V'_1\cup \ldots \cup V'_r \cup V'_{r+1}$ such that for $i:
1\leq i\leq r$, $(v,m_1,\ldots,m_n) \in V'_i$ iff $v\in
V_i$. And $(v,m_1,\ldots,m_n) \in V'_{r+1}$ iff $v\in V_{r+1}\cup \ldots
\cup V_n$. Let there be $r+1$ players such that the vertex set $V_i$
belongs to player $i$. Thus we introduce a dummy player, the $r+1$th
player, who owns all the vertices $(v,m_1,\ldots,m_n)\in V'$ such that
$v$ was originally an imitator vertex in $V$. By construction, we know
that every vertex $(v,m_1,\ldots,m_n)\in V'_{r+1}$ has an unique
outgoing edge $(v,m_1,\ldots,m_n)\step{a}(v',m'_1,\ldots,m'_n)$. Thus
the dummy player $r+1$ has no choice but to play this edge always. He
has a unique strategy in the arena $\game'$: at every vertex of $V'_{r+1}$,
play the unique outgoing edge.

\item Lift the preference orders of the players 1 to $r$ to
  subsets of 
  $V'$ as follows. A subset $W$ of $V'$ corresponds to the Muller
  set $F(W) = \{v\ \mid\ (v,m_{r+1},\ldots,m_n)\in W\}$ of
  $\game$. For every player $i: 1\leq i\leq r$, for $W, W'
  \subseteq V'$, $W \preceq'_i W'$ if and only if $F(W) \preceq_i
  F(W')$.

Since the player $r+1$ has a unique strategy and plays it always, his
preference ordering doesn't matter in the game. However, for
consistency, we assign the preference of an arbitrary imitator (say
imitator $n$) in the
game $(\game, v_0, \prec_1,\ldots,\prec_n)$ to the $r+1$th player in
the game $(\game', v'_0, \prec'_1,\ldots,\prec'_{r+1})$. That is, for $W, W'
  \subseteq V'$, $W \preceq'_{r+1} W'$ if and only if $F(W) \preceq_n
  F(W')$.
\end{enumerate}

The game $(\game', v'_0,\prec'_1,\ldots,\prec'_{r+1})$ is a turn based
game with $r+1$ players (the
optimisers and the dummy) such that each player $i$ has a preference ordering
$\preceq'_i$ over the Muller sets of $V'$. Such a game was called a
generalised Muller game in 
\cite{PS09}.

Let $L$ be the set

$$L=\{l \in (V'\cup\{\sharp\})^{|V'|+1}\ \mid\ |l|_\sharp = 1 \land
\forall v\in V'\ (|l|_v = 1)\}$$
where $|l|_v$ denotes the number of
occurences of $v$ in $l$. We have

\begin{theorem}[\cite{PS09}]
The game $(\game',v'_0, \prec'_1,\ldots,\prec'_{r+1})$ has a Nash
equilibrium in bounded memory 
strategies, the memory being $L$.
\end{theorem}

Now let
$\bar{\sigma'} = (\sigma'_1,\ldots,\sigma'_r,\sigma'_{r+1})$ be a Nash
equilibrium tuple 
for $r+1$ players in the game $(\game',v'_0,
\prec'_1,\ldots,\prec'_{r+1})$. We now construct a bounded
memory imitation equilibrium tuple $\bar{\sigma}$ for the $r$
optimisers in the game
$(\game,v_0, \prec_1,\ldots,\prec_n)$.

For the optimiser $i: 1\leq i\leq r$, let $\sigma'_i =
(L,\delta',g',l'_I)$. Define $\sigma_i = (M,\delta,g,l_I)$ to a
bounded memory strategy in the game
$(\game,v_0, \prec_1,\ldots,\prec_n)$ as
\begin{itemize}
\item $M= M_{r+1}\times\ldots\times M_n\times L$ where $M_i,\ r+1\leq
  i\leq n$ is the
  memory of strategy $\tau_i$ of imitator $i$. 
\item $\delta: A \times M \rightarrow M$ such that $\delta(a,\langle
  m_{r+1},\ldots, m_n,l\rangle) =
  \langle m'_{r+1},\ldots m'_n,\delta'(a,l)\rangle$ where $m'_i =
  \delta_i(a,m_i),$ $r+1\leq i\leq n$ such that $\delta_i$ is the
  memory update of
  strategy $\tau_i$.
\item $g: V \times M \rightarrow A$ such that $g(v,\langle
  m_{r+1},\ldots, m_n, l\rangle) = g'(\langle v,m_{r+1},\ldots,
  m_n\rangle, l)$.
\item $l_I = \langle m^{r+1}_I,\ldots m^n_I,l'_I\rangle$ where
  $m^i_I,\ r+1\leq i\leq n$
  is the initial memory of strategy $\tau_i$.
\end{itemize}

We then have:
\begin{theorem}
  $\bar{\sigma} = (\sigma_1,\ldots,\sigma_r)$ is
  an imitation equilibrium in 
  $(\game,v_0, \prec_1,\ldots,\prec_n)$.
\end{theorem}

\begin{proof} Suppose not and suppose player $i$ has an incentive to
  deviate to a strategy $\mu$ in $(\game,v_0,
  \prec_1,\ldots,\prec_n)$. Let $u \in A^\omega$ be the
  unique play consistent with the tuple
  $\bar{\sigma}$ where the imitators stick to
  their strategy tuple $(\tau_{r+1}, \ldots,\tau_n)$. Let $u' \in
  A^\omega$ be the 
  unique play consistent with the tuple
  $(\bar{\sigma}_{-i},\mu)$ (that is when player $i$ has
  deviated to the strategy $\mu$) where again the imitators stick to
  their strategy tuple $(\tau_{r+1}, \ldots,\tau_n)$. Let $l$ be
  the first index such that $u(l) \neq u'(l)$. Then, $v_0[u_{l-1}] \in
  V_i$, (where $u_{l-1}$ is the length $l-1$ prefix
  of $u$). That is, the vertex $v_0[u_{l-1}]$ belongs to optimiser $i$
  since everyone else sticks to her strategy.

Now consider what happens in the game
$(\game',v'_0,\prec'_{r+1},\ldots,\prec'_n)$ when all the optimisers
except
$i$ play the strategies
$\sigma'_1,\ldots,\sigma'_{i-1},\ldots,\sigma'_{i+1},\ldots,\sigma'_r$ and
the imitators stick to their strategy tuple $(\tau_{r+1},
\ldots,\tau_n)$. If the optimiser $i$ mimicks strategy $\mu$ for
$l-1$ moves in the game then the play is exactly $u_{l-1}$ and reaches
a vertex $(v,m_{r+1},\ldots,m_n)\in V'_i$ where $v=v_0[u_{l-1}]$. By
construction of the product, all the actions enabled at $v$ in the
arena $\game$
are also enabled in the arena $\game'$. Hence the optimiser $i$ can play
$u(l)$. By similar arguments, optimiser $i$ can mimick the strategy
$\mu$ in the arena $\game'$ forever. 

Thus by mimicking $\tau$ in the
game $(\game',v'_0,\prec'_{r+1},\ldots,\prec'_n)$, the optimiser 
$i$ can force a more preferable Muller set. But this contradicts the
fact that $\bar{\sigma}'$ is an equilibrium tuple in the game
$(\game',v'_0,\prec'_{r+1},\ldots,\prec'_n)$.
\end{proof}

\subsection{Stability}
Finally, we adress the questions asked in Section
\ref{sec1}. Given a game $(\game,v_0,\prec_1,\ldots,\prec_n)$ with
optimisers and imitators
where the optimisers play bounded memory strategies and the imitators
play imitative strategies specified by $k$ finite state transducers we
wish to find out:
\begin{itemize}
\item If a certain stongly connected component $W$ of $\game$ is where
  the play eventually settles down to.
\item What subtypes eventually survive.
\item How worse-off is imitator $i$ from an equilibrium outcome.
\end{itemize}

We have the following theorem:
\begin{theorem}\label{thm1}
  Let $(\game,v_0,\prec_1,\ldots,\prec_n)$ be a game with $n$ players
  where the first
  $r$ are optimisers playing bounded memory strategies $\sigma_1,\ldots,
  \sigma_r$ and the rest $n-r$ are imitators playing imitative
  strategies $\tau_{r+1},\ldots,\tau_n$ where every such strategy
  is among $k$ different types. Let $W$ be a strongly
  connected component of $\game$. The following questions are decidable:
  \begin{enumerate}
  \item[(i)] Does the game eventually settle down to $W$?
  \item[(ii)] What subtypes of the $k$ types eventually survive?
  \item[(iii)] How worse-off is imitator $i$ from an equilibrium outcome?
  \end{enumerate}
\end{theorem}

\begin{proof}
  Construct the arena
  $(\game',v'_0)=\game\times\fst_{\sigma_1}\times\ldots\times\fst_{\sigma_r}\times
  \R_{\tau_{r+1}}\times\ldots\times\R_{\tau_n}$.

  \begin{enumerate}
  \item[(i)] For the strongly connected component $S$ in
    $(\game',v'_0)$ that is reachable from $v'_0$, let $S$ be subgraph
    induced by the set $\{v\ \mid\ 
    (v,m_1,\ldots, m_n) \in S'\}$. Collapse the vertices of $S$ that
    have the same name and call the resulting graph $S''$. Check if
    $S''$ is the same as $W$ and output YES if so.
  \item[(ii)] For the strongly connected component $S$ in
    $(\game',v'_0)$ that is reachable from $v'_0$ do the
    following:
    \begin{itemize}
    \item For $i:\ r+1\leq i\leq n$ take the restriction of $S$ to
      the $i$th component for every
      $(v,m_1,\ldots,m_n)\in S$. Let $S_i$
      denote this restriction.
    \item Collapse vertices with the same name in $S_i$. Let $S'_i$ be
      this new graph.
    \item Check if $S'_i$ is a subtype of $\sigma_i$. If so output $S'_i$.
    \end{itemize}
  \item[(iii)] Compute a Nash equilibrium $\bar{\mu}$ of the game
    $(\game,v_0, \prec_1,\ldots,\prec_n)$
    using the procedure described in \cite{PS09}. Let $S'$ be the
    reachable strongly connected component of the arena
    $(\game',v'_0)$. Restrict $S'$ to the first component and call it
    $S$. Let $F=\mathit{occ}(S)$. Compare $F$ with
    $\inf(\rho_{\bar{\mu}})$ according to the preference ordering
    $\preceq_i$ of imitator $i$.
  \end{enumerate}
\end{proof}

\subsection{An Example}
Let us look at an example illustrating the concepts of the previous
section. Consider 3 firms A, B and C. Each firm has a choice of
producing 2 products, product $a$ or product $b$ repeatedly, i.e.,
potentially infinitely often. In every batch each of them can decide
to produce either of the products. 

Now firm A is a large firm with
all the technical knowhow and infrastructure and it can change between
its choice of production in consecutive batches without much
increase in cost. On the other hand, the firms B and C are small. For
either of them, if in any successive batch it decides to change from
producing $a$ to $b$ or vice-versa, there is a high cost incurred in
setting up the necessary infrastructure. Whereas, if it sticks to the
product of the previous batch, the infrastructure cost is
negligible. Thus in the case where it switches between products in
consecutive batches, it is forced to set the price of its product
high. This actually favours firm A as it can always set its product at
a reasonable price since it is indifferent between producing either of
the two products in any batch.

The demand in the market for $a$ and $b$ keeps changing. Firm A being
the bigger firm has the resources and knowhow to analyse the market
and anticipate the current demand and then produce $a$ or $b$
accordingly. Also assume that firm A is the first to put its product out
in the market. Thus it is tempting for firms B and C to imitate A. But
in doing so they run the risk of setting the prices of their products
too high and incurring a loss. 

\begin{figure}
\centering
\includegraphics[scale=0.6]{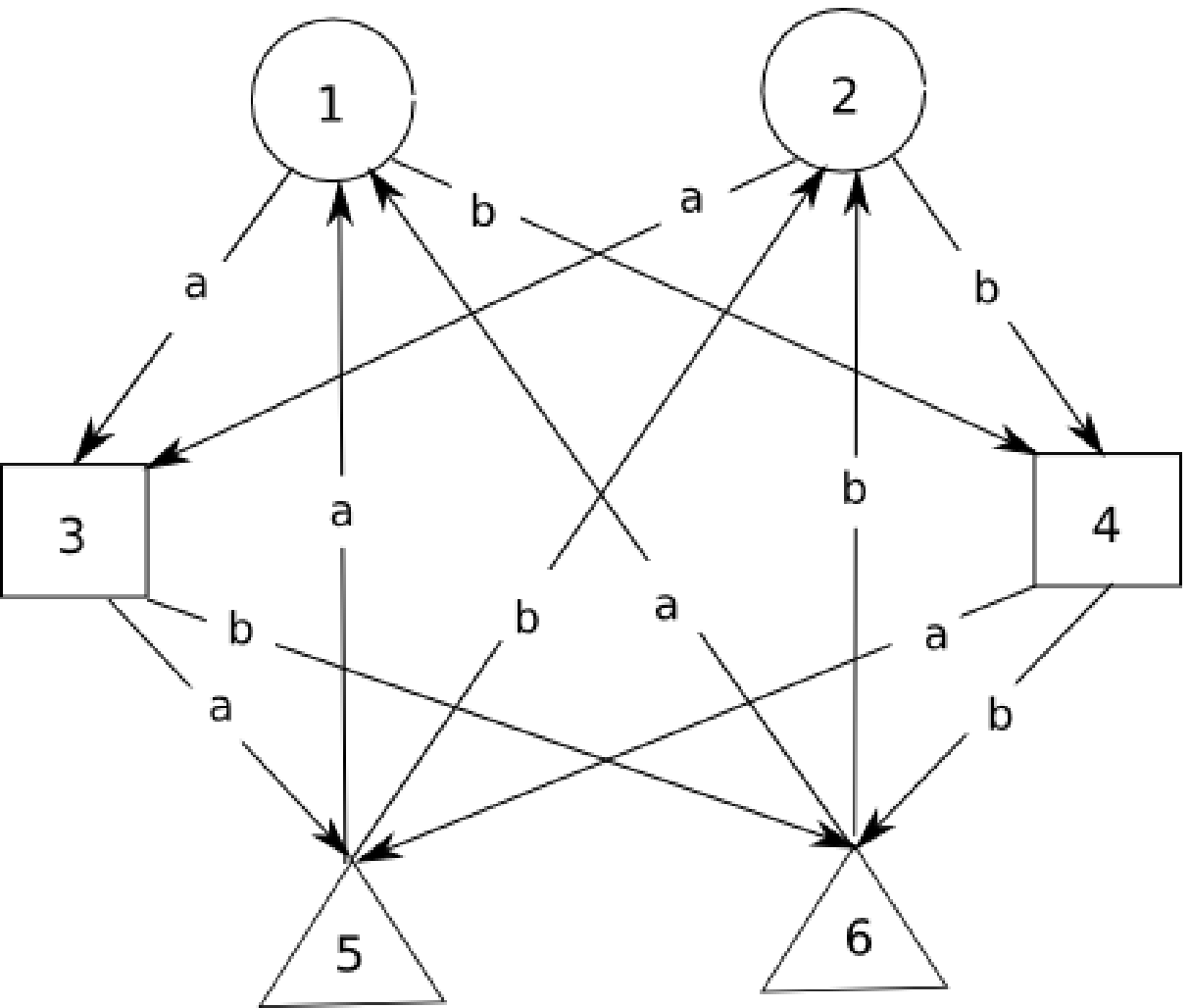}
\label{fig:figure1}
\caption{The arena $\game$}
\end{figure}

We model this situation in the form of the arena $\game$ shown in
Figure 2 where the nodes of firm A, B and C are denoted as $\bigcirc$,
$\Box$ and $\triangle$ respectively. The preferences of each of the
firms for the
relevant connected components when the market demand is low are given as:
$$\{1,2,3,4,5,6\} >_A X,\ \text{for } X\subsetneq \{1,2,3,4,5,6\}$$
$$\{1,3,5\} >_B \{1,4,5\} >_B \{1,3,5,4\} >_B \{2,3,6,4\} >_B Y,\
\text{for any other } Y\subsetneq \{1,2,3,4,5,6\}$$
$$\{1,3,5\} >_C \{1,4,5\} >_C  \{2,3,6,4\} >_C \{1,3,5,4\} >_C Z,\
\text{for any other } Z\subsetneq \{1,2,3,4,5,6\}$$
Thus firm A prefers the larger set $\{1,2,3,4,5\}$ to the smaller ones
while B and C prefer the smaller sets. But when the market demand is
high their preferences are given as:
$$\{1,2,3,4,5,6\} >_i X,\ \text{for } X\subsetneq
\{1,2,3,4,5,6\}\text{ and }i\in \{A, B, C\} $$
That is, all of them prefer the larger set.

Now if A produces $a$ and $b$ in alternate batches and $B$ and $C$
imitate A, then we end up in the component $\{1,2,3,4,5,6\}$ which is
profitable for A but less so for B and C when the market demand is
not so high. But when the demand is high, the component
$\{1,2,3,4,5,6\}$ is quite profitable even for B and C and thus in
this case, imitation is a viable strategy for them.


\section{Discussion}\label{sec5}
The model that we have presented here is far from definitive, but we
see these results as 
early reports in a larger programme of studying games with player
types. The model 
requires modification and refinement in many directions, being
addressed in related 
on-going work. In games with large number of players, outcomes are
typically associated 
not with player profiles but with distribution of types in the
population. Imitation 
crucially affects such dynamics. Our model can be easily modified to incorporate
distributions but the analysis is considerably more
complicated. Further, it is natural 
to consider this model in the context of repeated normal form games,
but in such contexts 
almost-sure winning randomized strategies are more natural. A more
critical notion 
required is that of type based reduction of games, so that analysis of
large games can be 
reduced to that of interaction between player types.

\subsection*{Acknowledgement} We thank the anonymous
referees for their helpful comments and suggestions. The second author
thanks NIAS (http://nias.knaw.nl) for support when he was working on
this paper.


\end{document}